\documentstyle[12pt]{article}
\begin{document}
\baselineskip = 20pt

\title{THE PARTITION FUNCTION FOR AN ANYON-LIKE OSCILLATOR}

\author{
H. Boschi-Filho$^{\star}$, C. Farina\\
\\
Instituto de F\'\i sica - Universidade Federal do Rio de Janeiro \\
Cidade Universit\'aria - Ilha do fund\~ao - Caixa Postal 68528 \\
21945-970 Rio de Janeiro, BRAZIL.\\
\\
\\
A. de Souza Dutra$^\ast$\\
\\
Universidade Estadual Paulista - Campus de Guaratinguet\'a\\
Departamento de F\'\i sica e Qu\'\i mica, \\
Av. Dr. Ariberto Pereira da Cunha, 333\\
CEP 12500, Guaratinguet\'a - SP - BRAZIL.}

\bigskip
\maketitle
\begin{abstract}
We compute the partition function of an anyon-like harmonic
oscillator. The well known results for both the bosonic and fermionic
oscillators are then reobtained as particular cases as ours. The
technique we employ is a non-relativistic version of the Green
function method used in the computation of one-loop effective actions
of quantum field theory.
\end{abstract}
\bigskip
\vfill
\noindent $^\star${e-mail: BOSCHI@VMS1.NCE.UFRJ.BR}
\par
\noindent $^\ast${e-mail:
DUTRA@GRT$\emptyset\emptyset\emptyset$.UESP.ANSP.BR}
\par

\pagebreak

Partition functions give the statistical behavior of a system of
particles in thermal equilibrium with each other and with a thermal
bath.  In general, the search for these functions is not an easy task
although for some particular systems there are well known results.
Naturally, many techniques are available for doing this work.  In
particular, Gibbons \cite{Gibbons} used the fact that the partition
functions for the bosonic and fermionic oscillators could be written
as determinants of the relevant operator, with periodic and
antiperiodic boundary conditions, respectively, in order to compute
them through the $\zeta$-function method.

Specifically speaking, it follows directly from the definition of a
partition function that for any bosonic system we can write
\begin{eqnarray}
Z^B(\beta)&=& Tr \; e^{-\beta H}\nonumber\\
&=& \int dx_0 <x_0\vert e^{-\beta H}\vert x_0>\nonumber\\
&=&\int dx_0 K(x_0,x_0;\tau=-i\hbar\beta),
\end{eqnarray}
where $K(x,y;\tau)$ is the usual Feynman propagator. For the bosonic
(harmonic) oscillator, we substitute its path integral representation
and get
\begin{eqnarray}
Z^B(\beta)&=& \int dx_0 \int_{{x(0)=x_0}\atop{x(\beta)=x_0}}  [Dx]
e^{-{1\over 2}
\int_0^\beta x(\tau)(\omega^2-\partial^2)x(\tau) d\tau}\nonumber\\
&=& \int_{x(0)=x(\beta)} [Dx] \, e^{-{1\over 2}
\int_0^\beta x(\tau)(\omega^2-\partial^2)x(\tau) d\tau}\nonumber\\
&=& {\det}^{-1/2} (\omega^2-\partial^2)\vert_{periodic}
\end{eqnarray}

Analogously, using standard Grassmann variables, it can be shown that
the partition function for the (second order) fermionic oscillator is
given by \cite{Gibbons},\cite{Berezin}

\begin{equation}
Z^F(\beta)= {\det}^1(\omega^2-\partial^2)\vert_{antiperiodic}
\end{equation}

Here, we will in fact generalize these results by making analogous
calculations, but this time we shall impose a generalized boundary
condition which contains as particular cases the periodic and
antiperiodic conditions discussed in Gibbon's paper.
Besides, we shall use an alternative technique which is the non
relativistic version of the Green function method for computing
effective actions in quantum field theory \cite{Schwinger},\cite{Reuter}.

Suppose, then, we want to compute the following determinant

\begin{equation}\label{det}
\exp[\Gamma_s^\theta
(\omega)]={\det}^{s}\left(\omega^2+\partial_t^2\right)_\theta
\equiv{\det}^{s}\left( L\right)_\theta,
\end{equation}

\noindent
where $L$ acts on functions that satisfy some given boundary
condition specified by the label $\theta$. The power of
the determinant given by the parameter $s$ is left completely
arbitrary to take into account the cases that interpolate between
the fermionic and the bosonic oscillators. Hence, playing with the boundary
condition and the parameter $s$ we can pass continuously from the
bosonic to the fermionic oscillator. That is why we refer
to this determinant as the partition function of an \lq\lq
anyon-like"  oscillator.

Suggested by the Green function method usually employed in quantum
field theory, we write
\begin{eqnarray}
{\partial\over\partial\omega}\Gamma_s^\theta (\omega) &=& 2s\omega
Tr\left( L^{-1}\right)_\theta \nonumber \\
&=& 2s\omega\int_0^\tau
G_\omega^\theta (t,t)dt\,,
\end{eqnarray}
where the Green function $G^\theta_\omega(t,t')$ satisfies
\begin{equation}\label{eq}
\left(\omega^2+\partial_t^2\right)G_\omega^\theta (t,t')=\delta(t-t'),
\end{equation}
as well as some boundary condition (to be given in a moment).
Integrating the above equation we obtain

\begin{equation}\label{Gamma}
\Gamma_s^\theta (\omega)-\Gamma_s^\theta(0)=2s\int_0^\omega
d\omega' \omega'\int_0^\tau dt G_{\omega'}^\theta(t,t)\,.
\end{equation}

Since our purpose here is to use a generalized boundary condition in
the sense that the periodic and antiperiodic cases will appear as
particular cases, and in order to make connection with
the behavior of correlation functions of anyon-like systems, it is
natural to impose the following $\theta$-dependent condition
\begin{equation}\label{cond}
G_\omega^\theta(t+\tau,t')= e^{-i\theta}G_\omega^\theta(t,t').
\end{equation}

\noindent It is clear that this boundary condition becomes periodic
for $\theta=0$ and antiperiodic for $\theta=\pi$.  Depending on these
conditions and the value of the parameter $s$, which can be thought
as a ``statistical'' parameter, this determinant will be mapped into
different partition functions.  As these particular cases are related
to bosonic and fermionic systems, this condition of general
periodicity could, in principle, be related to particles whose
statistics interpolates bosons and fermions, {\it i. e.}, anyons
\cite{Wilczek}.

It is straightforward to construct the Green function $G_\omega^\theta
(t-t')$. Using basically the same technique that Kleinert
\cite{Kleinert} employed for the simpler cases of periodic and
antiperiodic boundary conditions, it can be shown that (see the
Appendix)
\begin{equation}\label{G}
G_\omega^\theta(t-t')=
{e^{-i\theta/2}\over 4\omega}\biggl[{e^{i\omega(t-t'-\tau/2)}\over
\sin({\omega\tau+\theta\over2})} + {e^{-i\omega(t-t'-\tau/2)}\over
\sin({\omega\tau-\theta\over2})} \biggr]\;\; ;\;\; t-t'\
\epsilon\;[0,\tau)
\end{equation}

Substituting this Green function in Eq. (\ref{Gamma}), for the interval
$[0,\tau)$ and with $t=t^\prime$, we have
\begin{eqnarray}
\Gamma_s^\theta(\omega)-\Gamma_s^\theta(0) &=&
2s\int_0^\tau dt\int_0^\omega
d\omega^\prime \omega^\prime\biggl\{{e^{-i\theta/2}\over
4\omega'}\biggl[{e^{-i\omega'\tau/2}\over
\sin({\omega'\tau+\theta\over2})} + {e^{i\omega'\tau/2}\over
\sin({\omega'\tau-\theta\over2})} \biggr]\biggr\}\nonumber \\
&=&
\ln\left\{ e^{i\theta} \left[-1+e^{-i(\theta+\omega\tau)}\right]
\left[1-e^{-i(\theta-\omega\tau)}\right]\right\}^s. \label{z}
\end{eqnarray}

\noindent Recalling Eq.(\ref{det}) we see that the exponential of
$\Gamma_s^\theta(\omega)$ is the desired determinant. Identifying
$\tau=-i\beta$ ($\hbar=1$), taking $\theta=0$ (periodic boundary
condition) and $s=-1/2$, this determinant reduces, apart from a
constant factor  $\exp[\Gamma_s^\theta(0)]$, which hereafter we call
$C$, to the partition function for a bosonic oscillator
\cite{Gibbons}, \cite{Kleinert}, \cite{Das}, namely,
\begin{equation}
Z^{B}(\beta) = \exp\left[\Gamma_{-1/2}^{0}
(\omega)\right]
= {C\over 2 \sinh ({\omega\beta\over2})},
\end{equation}

\noindent Note that any thermodynamical quantity which can be
obtained form the partition function does not depend on $C$.

Analogously, for a fermionic oscillator, we just make $\theta=\pi$
(antiperiodic boundary condition) and $s=+1$, so that
$\exp\left[\Gamma_s^\theta(\omega)\right]$ reduces to the following
partition function
\begin{equation}
Z^{F(2)}(\beta) = \exp\left[\Gamma_{+1}^{\pi}
(\omega)\right] = 4 C \cosh^2({\omega\beta\over2}).
\end{equation}

\noindent This result corresponds to the partition function for
a second order fermionic oscillator. One can check it by calculating
explicitly the eigenvalues for the Finkelstein-Villasante Grassmann
oscillator (with N=2)  \cite{Finkelstein} and then finding its
partition function by summing the trace of $\exp(-\beta E_n)$.
This result differs from the one given by Gibbons \cite{Gibbons} in
the quadratic power of $\cosh({\omega\beta\over2})$, once he considered an
equivalent linear Grassmann oscillator opposed to the quadratic case
discussed here. This linear case can also be obtained from our
discussion if we take from the very beginning the determinant of $L^{1/2}$
instead of $L$, so that we find
\begin{equation}
Z^{F(1)}(\beta) = \exp\left[\Gamma_{+1}^{\pi}
(\omega)\right]_{Linear} = 2 C \cosh ({\omega\beta\over2}),
\end{equation}

\noindent which is the well known partition function for the linear
fermionic oscillator \cite{Gibbons},
\cite{Kleinert}, \cite{Das}.

For the general case, the partition function reads
\begin{eqnarray} \label{Z}
Z_s^{\theta}(\beta) & \equiv &
\exp\left[\Gamma_{s}^{\theta} (\omega)\right]  \nonumber \\ & = &
 \left[ e^{-i\theta} \left(-1+e^{-i(\theta+\omega\tau)} \right)
\left(1-e^{-i(\theta-\omega\tau)} \right )\right]^s \nonumber \\
& = & 4^s  \left[\cosh^2 {\omega\beta\over 2}
- \cos^2 {\theta\over 2} \right]^s \label{Ztheta}\,,
\end{eqnarray}

\noindent where for simplicity we put $C=1$. Naturally, the above
calculated partition functions are particular cases of Eq. (\ref{Z}).

Note that, in Eq. (\ref{Z}), we left the statistics parameter $s$
free.  In fact, it may be a function of the periodicity parameter
$\theta$, interpolating between $s(\theta=0)=-1/2$ (bosonic case) and
$s(\theta=\pi)=+1$ (fermionic case), as for example $s(\theta)=-1/2
+3/2 f(\theta)$, where $f(\theta)$ may be a function which satisfies
$f(0)=0$ and $f(\pi)=+1$.  We wonder if this factor can be obtained
from the functional integration of a generalized variable, with
arbitrary commutation relation, interpolating the cases of bosonic
(c-number) and Grassmannian variables, as a kind of a q-deformed
calculation \cite{Lerda},\cite{Zeluis}.

Another possible interpretation for this generalized partition
function is in relation with parasystems \cite{Ohnuki}, where there
is a parameter for which convenient limits reproduce the bosonic and
fermionic oscillators \cite{Das2}. However, the connection between these
systems with the present calculations (if any) seems to be
non-trivial and deserves further study. This will be discussed
elsewhere.

\bigskip
\bigskip

\bigskip
\bigskip

One of the authors (C.F.) would like to thank to M.Asorey and
A.J.Segu\'\i- Santonja for enlightening discussions and to the
Theoretical Physics Department of the University of Zaragoza for
hospitality during his stay in Zaragoza, where part of this work was
done.  The authors would like also to acknowledge A. Das for helpful
discussions. This work was partially supported by MEC-CAICYT and
Conselho Nacional de Desenvolvimento Cient\'\i fico e Tecnol\'ogico
(CNPq).

\bigskip
\bigskip

\bigskip
\centerline{\bf Appendix}
\bigskip

In this Appendix we construct the Green function Eq.(\ref{G}).  The
following discussion is similar to that found in Kleinert
\cite{Kleinert}. In fact, we generalize Kleinert's calculations for
the cases where the boundary conditions are neither periodic and
antiperiodic, but anyon-like conditions.

The spectral representation for $G_\omega^\theta(t)$ is given by
\begin{eqnarray}
G_\omega^\theta(t)&=&{1\over\tau}\sum_{m=-\infty}^\infty
{e^{-i{\omega_m^\theta} t}\over \omega^2-{\omega_m^\theta}^2}
\nonumber \\
&=&{1\over\tau}\sum_{m=-\infty}^\infty\biggl\{{e^{-i{\omega_m^\theta} t}
\over 2i\omega}
\biggl[{i\over \omega-{\omega_m^\theta}}+{i\over \omega+{\omega_m^\theta}}
\bigg]
\biggr\}\nonumber \\
&=&{1\over 2\omega
i}\biggl\{G_-^\theta(t)+G_+^\theta(t)\biggr\},\label{spectral}
\end{eqnarray}

\noindent
where ${\omega_m^\theta}=(2\pi m+\theta)/\tau$ and we identified the
spectral representation of $G_\pm^\theta(t)$, which are respectively
the Green functions associated to the first order operators
$(i\partial_t\pm\omega)$ also satisfying the generalized periodic
boundary condition, Eq. (\ref{cond}).

Using the Poisson summation formula \cite{GR}
\begin{equation}
\sum_{m=-\infty}^\infty
f(m)=\int_{-\infty}^\infty d\mu
\sum_{n=-\infty}^\infty e^{2\pi i\mu n}f(\mu),
\end{equation}

\noindent
we can write for $G_\pm^\theta(t)$
\begin{eqnarray}
G_\pm^\theta(t)&=&{i\over\tau}\sum_{m=-\infty}^\infty
{e^{-i{\omega_m^\theta} t}\over \omega\pm{\omega_m^\theta}}\nonumber \\
&=&\sum_{n=-\infty}^\infty
\int_{-\infty}^\infty {d\omega'\over 2\pi}
e^{-i(\omega't-n\omega'\tau+n\theta)}\biggl({i\over\omega\pm\omega'}
\biggr).\label{G+-}
\end{eqnarray}

\noindent Identifying
\begin{equation}
G_\pm(t)=\int_{-\infty}^\infty {d\omega'\over
2\pi}e^{-i\omega' t}
\biggl({i\over \omega\pm\omega'}\biggr)
\end{equation}

\noindent
as the Green functions associated with the first order operators
$(i\partial_t\pm\omega)$, but this time valid for an infinite time
interval, we may cast (\ref{G+-}) into the form
\begin{equation}\label{7}
G_\pm^\theta(t)=\sum_{n=-\infty}^\infty e^{-in\theta} G_\pm(t-n\tau).
\end{equation}

\noindent Hence, to obtain $G_\pm^\theta(t)$, we need first to
compute $G_\pm(t)$. By residue calculations this can be made after
setting $\omega\rightarrow\omega-i\eta$. With this prescription, it
can be easily shown that
\begin{equation}\label{8}
G_\pm (t)=-\Theta(\mp  t) e^{\pm  i\omega t}\,,
\end{equation}

\noindent where $\Theta(t)$ is the usual Heaviside step function.
Substituting (\ref{8}) into (\ref{7}), we get
\begin{equation}\label{9}
G_\pm ^\theta(t)= - \sum_{n=-\infty}^\infty e^{-in\theta} \Theta(\mp
t\pm n\tau) e^{\pm  i\omega(t-n\tau)}.
\end{equation}

\noindent Since this expression has a period $\tau$, we can restrict
ourselves to the interval $t\epsilon[0,\tau)$. Hence, the sum
appearing in the r.h.s. of (\ref{9}) can be easily done, yielding for
$G_+^\theta(t)$
\begin{eqnarray}
G_+^\theta(t)&=&-e^{i\omega t}\sum_{n=1}^{+\infty} e^{-i n
(\omega\tau +\theta)}\nonumber \\
&=&-e^{i\omega t}\biggl\{e^{-i(\omega\tau+\theta)}+
e^{-2i(\omega\tau+\theta)}+...\biggr\}\nonumber \\
&=&-{e^{i\omega t} e^{-i(\omega\tau+\theta)}\over
1-e^{-i(\omega\tau+\theta)}}.\label{10a}
\end{eqnarray}

\noindent Analogously, for $G_-^\theta(t)$ we have
\begin{eqnarray}
G_-^\theta(t) &=& -e^{-i\omega t}\sum_{n=0}^{-\infty} e^{-i n
(\omega\tau -\theta)}\nonumber \\
&=&-e^{-i\omega t}\biggl\{1+e^{-i(\omega\tau-\theta)}+
e^{-2i(\omega\tau-\theta)}+...\biggr\}\nonumber \\
&=&-{e^{-i\omega t}\over
1-e^{-i(\omega\tau-\theta)}}.\label{10b}
\end{eqnarray}

\noindent
Of course, outside this interval, the result can be obtained by
periodicity.  Substituting (\ref{10a}) and (\ref{10b}) into
(\ref{spectral}) and rewriting $t$ as $t-t'$, we finally obtain
\begin{equation}
G_\omega^\theta(t-t')=
{e^{-i\theta/2}\over 4\omega}\biggl[{e^{i\omega(t-t'-\tau/2)}\over
\sin({\omega\tau+\theta\over2})} + {e^{-i\omega(t-t'-\tau/2)}\over
\sin({\omega\tau-\theta\over2})} \biggr]\;\; ;\;\; t-t'\
\epsilon\;[0,\tau)
\end{equation}

\noindent This Green function satisfies Eqs. (\ref{eq}) and
(\ref{cond}).  This formula generalizes the previous known particular
cases of periodic $(\theta=0)$ and anti-periodic $(\theta=\pi)$
boundary conditions, for which we find the well known results in
accordance with the literature \cite{Kleinert}.

\vfill\eject


\begin{thebibliography}{99}

\bibitem{Gibbons} G.W.Gibbons, Phys. Lett. {\bf A60}, 385 (1977).

\bibitem{Berezin} F.A. Berezin, {\it The method of second
quantization} (Academic Press, New York, 1966).

\bibitem{Finkelstein} R. Finkelstein and M. Villasante, Phys. Rev.
{\bf D 33} (1986) 1666.

\bibitem{Schwinger} J. Schwinger, Phys. Rev. {\bf 82}, 664 (1951).

\bibitem{Reuter} W. Dittrich and M. Reuter, {\it Lecture
Notes in Physics: Effective Lagrangians in Quantum Electrodynamics}
(Springer-Verlag, Berlin, 1984), Vol.{\bf 220}.

\bibitem{Wilczek} F. Wilczek, {\it Fractional statistics and anyon
superconductivity}, (ed.) World Scientific, Singapore, 1990.

\bibitem{Kleinert} H. Kleinert, {\it Path Integrals in Quantum Mechanics,
Statistics, and Polymer \break Physics}, World Scientific Publishing,
Singapore (1990).

\bibitem{Das} A. Das, {\it Quantum field theory, a path integral
approach}, World Scientific, 1990.

\bibitem{Lerda} A. Lerda and S. Sciuto, Nucl. Phys. B {\bf 401}
(1993) 613.

\bibitem {Zeluis} J.L. Matheus-Valle and M.A.R. Monteiro, Mod.
Phys. Lett. A {\bf 9} (1994) 945.

\bibitem{Ohnuki} Y. Ohnuki and S. Kamefuchi, {\it Quantum field theory and
parastatistics}, (Univers. of Tokyo Press, Springer-Verlag, Berlin, 1982).

\bibitem{Das2} S.N. Biswas and A. Das, Mod. Phys. Lett. A {\bf 3}
(1988) 549.

\bibitem{GR} I.S.Gradshteyn and I.M.Ryzhik, {\it Table of Integrals,
Series, and Products} (Academic, New York, 1980).



\end{thebibliography}
\end{document}